\documentclass{epl}
\usepackage{amssymb,amsmath,amsfonts}
\usepackage{epsfig}

\def\be{\begin{equation}}
\def\ee{\end{equation}}
\def\bea{\begin{eqnarray}}
\def\eea{\end{eqnarray}}

\def\vs{\vec{S}}
\def\vh{\vec{h}}

\title{Driven Heisenberg Magnets: Nonequilibrium Criticality, Spatiotemporal
Chaos and Control}
\author{J. Das\inst{1,2}\thanks{Present Address:  Center for Stochastic Processes
in Science and Engineering and Department of Physics, Virginia Tech,
Blacksburg, VA 24061-0435, USA. E-mail:\email{jayajit@vt.edu}} 
\and  
M. Rao\inst{1}\thanks{E-mail: \email{madan@rri.res.in}}
\and S. Ramaswamy\inst{3}
\thanks{Also affiliated with Jawaharlal Nehru Centre for
Advanced Scientific Research,
Bangalore 560 064 India.  E-mail: \email{sriram@physics.iisc.ernet.in}}}
\institute{
  \inst{1} Raman Research Institute, C.V. Raman Avenue,
Sadashivanagar, Bangalore 560080, India \\
  \inst{2}Institute of Mathematical Sciences, Taramani, Chennai
600113, India \\
  \inst{3} Centre for Condensed-Matter Theory,
Department of Physics, Indian Institute of Science, Bangalore 560
080, India \\
}


\begin{document}

\maketitle

\begin{abstract}
We drive a $d$-dimensional Heisenberg magnet using an anisotropic current. The
continuum Langevin equation is analysed using a dynamical renormalization group
and numerical simulations. We discover a rich steady-state phase diagram,
including a critical point in a new nonequilibrium universality class, and a
spatiotemporally chaotic phase. The latter may be `controlled' in a robust manner
to target spatially periodic steady states with helical order.
\end{abstract}

How does an imposed steady current of heat or particles alter the dynamics of the 
isotropic magnet? To answer this question, we extend the equations of motion 
for the classical $O(3)$ Heisenberg model \cite{MAMAZ} to include the effects of 
a uniform current in one spatial direction while retaining isotropy in the 
order parameter space. The resulting model is a 
natural generalization of the driven diffusive models of \cite{DDLG} to the case of a 
3-component {\em axial}-vector order parameter and, as such, is an important step in 
the exploration of dynamic universality classes \cite{HH} far from equilibrium
\cite{UWERUIZ}. 
The form of the local molecular field in which spins precess in this driven state 
is strikingly different from that at equilibrium \cite{MAMAZ}, and is responsible for
all the remarkable phenomena we predict, including a novel nonequilibrium 
critical point and, in a certain parameter range, a type of turbulence.

Here are our results in brief: (i) Despite $O(3)$ invariance in the order-parameter 
space, the dynamics does not conserve magnetization; (ii) As a temperature-like 
parameter is lowered, 
the paramagnetic phase of the model approaches a nonequilibrium critical point in a 
new dynamic universality class; (iii) Below this critical point, in mean-field theory
without stochastic forcing, paramagnetism, ferromagnetism and helical order are all
linearly unstable; (iv) Numerical
studies in space dimension $d=1$ show spatiotemporal chaos in this last regime. This
chaos, when `controlled', is replaced by spatially periodic steady helical states.
These predictions should be testable in experiments on isotropic magnets carrying a 
steady particle or heat current, as well as in simulations of a magnetized lattice-gas 
which we discuss at the end of this Letter.   

To construct our equations of motion, recall that  
{\em at thermal equilibrium} at temperature $T$ the probability of spin configurations 
$\{\vs_i\}$ of a general nearest-neighbor Heisenberg chain 
with sites $i$ is $\propto \exp(-H/T)$, with an energy
function  
\be
\label{heis}
H = - \sum_i J_i \vs_i.\vs_{i+1}, 
\ee
where $J_i$ is the exchange coupling between $i$ and $i + 1$. 
A spin $\vs_i$ at $i$ precesses as $\dot{\vs}_i = \vs_i \times \vh_i$ where 
\be
\label{heq}
\vh_i = -{\partial H \over \partial \vs_i} = J_i\vs_{i+1} + J_{i-1}\vs_{i-1}
\ee
is the local molecular field. Replacing $J_i \to J(x)$ 
and $\vs_i \to \vs(x)$ in the continuum limit, yields \cite{radha} 
$\dot{\vs}(x) = J(x) \vs \times \partial^2_x \vs + (dJ/dx) \vs \times \partial_x \vs +
...$. For the physically reasonable case where $J$ varies {\em periodically} about a mean 
value $J_0$, this reduces for long wavelengths to $\dot{\vs}(x) = J_0 \vs \times
\partial^2_x \vs$, which is invariant under $x \to -x$, even if the 
$H$ is not. The dynamics conserves  $\sum_i\vs_i$ since it commutes with $H$.   

Now drive some background degrees of freedom in, say, the ${\bf \hat{x}}$ direction, 
{\em retaining isotropy in spin space}. These could be some mobile species -- particles, 
vacancies, heat -- or some nonconserved internal variables.  
Possible microscopic realizations are discussed towards the end 
of the paper. For now, note that the dynamics in this nonequilibrium state  
does not follow from an energy function, and must be constructed anew. 
If we average over these background variables,
their effect should be simply to modify the equations for 
the $\vs_i$ by allowing terms forbidden at {\em thermal} 
equilibrium. While many such terms are permitted, only two are {\em relevant}: 
(i) asymmetric exchange, i.e.,  
$\vh_i = J_+\vs_{i+1} + J_-\vs_{i-1}$ 
yielding a precession rate $g \vs \times \partial^2_x \vs + 
\lambda \vs \times \partial_x \vs$, with $\lambda \propto J_+ - J_-$ proportional 
to the driving rate;  
and (ii) nonconserving damping and noise. Both (i) and (ii) were ruled 
out \cite{MAMAZ} at thermal equilibrium only because the dynamics {\em had} 
to be generated by (\ref{heis}) and (\ref{heq}). 
Note that the $\lambda$ term, while
rotation-invariant in spin space, is not the divergence of a current \cite{KPZ}.    
The nonlinearity $\lambda {\vs}\times \partial_x{\vs}$ will thus {\em generate} 
\cite{NOTE,THESIS,BIG} 
nonconserving noise and damping terms even if these are not put in at the outset. 

For a general dimension $d \equiv d_{\perp}+1$, with anisotropic driving 
along one direction (${\|}$) only, the above arguments yield,  
to leading orders in a
gradient expansion, the generalized Langevin equation 
\begin{eqnarray}
\frac{\partial {\vs}}{\partial t} & =  & \bigg( r_{\|}\partial_{\|}^2 
+ r_{\perp}\nabla_{\perp}^2 \bigg) {\vs} - v
{\vs}-\frac{u}{6}({\vs}\cdot{\vs}){\vs}
-\lambda {\vs}\times \partial_{\|}{\vs} \nonumber \\
 &   &  +g_{\|}{\vs}\times \partial_{\|}^2{\vs}
+g_{\perp}{\vs}\times \nabla_{\perp}^2{\vs} + \vec{\eta}\, ,
\label{eq:dyeq}
\end{eqnarray}
where we have allowed for spatial anisotropy in the coefficient of the 
usual spin 
precession term. The Gaussian, zero-mean nonconserving
noise $\vec \eta$ satisfies  
$\langle \eta_{\alpha}({\bf x},t) \eta_{\beta}({\bf x}',t')\rangle = 
2B\,\delta_{\alpha \beta}\,\delta^d({\bf x}-{\bf x}') \delta(t-t')\,.$
 
In the {\em equilibrium, isotropic} limit, $\lambda = u = v \equiv 0$, $r_{\|} = r_{\perp} 
\equiv r$, the noise strength vanishes at zero wavenumber, and (\ref{eq:dyeq}) 
has a critical  
point where the renormalized $r \to 0$. In the {\em driven state}, since the dynamics and
noise are nonconserving, the critical point is $v = 0$, which in general takes place on
a curve in the temperature/driving-force plane. As the drive is taken to zero
there should be a crossover from nonequilibrium to equilibrium critical behavior. 
Our primary interest is in the behavior at a given
nonzero driving rate, for which it suffices to vary the temperature-like parameter 
$v$ in (\ref{eq:dyeq}), keeping the rest fixed (with $r_{\|}, \,  r_{\perp}, u > 0$). 
For $v>0$ (the paramagnetic phase) all correlations clearly decay on finite lengthscales
$\sim 1 / \sqrt{v}$  
and time-scales $\sim 1/v$, and nonlinearities are irrelevant. Let us focus
first on the nature of correlations on the critical surface $v = 0$. Here,   
we expect an anisotropic scaling form for the correlation function  
$C(x,t) \equiv \langle \vs  ({\bf x}+{\bf x}^{'},t+t^{'})\cdot{\vs} 
({\bf x}^{'},t^{'})\rangle$:  
$C({\bf x},t)=x_{\|}^{2\chi}
F (t/x_{\|}^z, x_{\perp}/x_{\|}^{\zeta})$,
where $F$ is a scaling function. In the {\em linear} approximation to (\ref{eq:dyeq}) 
the roughening, growth, and anisotropy exponents are respectively $\chi=1-d/2, \, z=2, 
\, {\rm and} \, \zeta=1$, and $F$ is analytic in its arguments. 

We now include the effect of the nonlinear terms in (\ref{eq:dyeq})
via a standard implementation of the dynamical 
renormalization group (DRG) \cite{MA} based on a perturbation
expansion in $\lambda$ and $u$. 
Rescaling  
$x_{\|}=bx_{\|}',\,{\bf x}_{\perp}=b^{\zeta}{\bf x}_{\perp}',\, t=b^zt'$
and ${\vs}=b^{\chi} {\vs}'$, where $b > 1$ is an arbitrary parameter, 
the coefficients in Eq.
(\ref{eq:dyeq}) transform as $r'_{\|}=b^{z-2}r_{\|}$,
$r'_{\perp}=b^{z-2\zeta}r_{\perp}$, $B'=b^{z-2\chi-\zeta (d-1)-1}B$,
$u'=b^{4-d}u$, $\lambda^{'}=b^{\chi+z-1}\lambda$, $g_{\|}^{'}
=b^{\chi+z-2}g_{\|}$ and $g_{\perp}^{'}=b^{\chi+z-2 \zeta}g_{\perp}$.
${\lambda}$ and $u$ are thus {\em relevant} 
for dimension $d < 4$, and   
$g_{\|}$ and $g_{\perp}$ are {\it irrelevant} for $d$ near $4$.  
%
In units where the ultraviolet cutoff is 1, $\lambda$ and $u$ enter the perturbation 
theory in the dimensionless combinations $\tau \equiv (1/2 \pi^3)\lambda^2\,B
/\sqrt{r_{\|}^3\,r_{\perp}^3}$
and $\kappa \equiv (1/2 \pi^3) u\,B /\sqrt{r_{\|} r_{\perp}^{3}}$, 
{\em At the critical point} $v = 0$, setting the
irrelevant $g_{\|}$ and $g_{\perp}$ to zero, for $d = 4 - \epsilon$, 
we find \cite{THESIS,BIG} to $O(\epsilon)$ the differential recursion relations 
\begin{eqnarray}
\frac {\partial {r}_{\|}}{\partial l} & = & {r}_{\|} (
z-2+\frac{\pi}{4} {\tau}), \nonumber \\
\frac{\partial {r}_{\perp}}{\partial l}  
& =  & {r}_{\perp} (
z-2\zeta+\frac{5\pi}{48}{\tau} ), \nonumber \\
\frac{\partial {B}}{\partial l} & = & {B} [
z-2\chi-\zeta(d-1)-1+\frac{\pi}{32} {\tau} ],
\nonumber \\
\frac{\partial {\tau}}{\partial \ell} & = &
{\tau}\, (\epsilon \zeta-\frac{35}{64}\pi {\tau}), 
\nonumber \\
\frac{\partial {\kappa}}{\partial \ell} & = &
{\kappa}\,(
\zeta\epsilon- \frac{11}{24}\,\pi\zeta{\kappa}-
\frac{\pi}{2}{\tau})+ \frac{27}{16}\pi\zeta\,{\tau}^2. 
\end{eqnarray}
Since we are working at $v=0$, we seek a fixed point that is {\em
stable} with 
respect to perturbations in the {\em remaining}
directions in the parameter space. 
For $\epsilon = 4 - d > 0$ we find the nontrivial stable fixed point 
${\tau}^*=64\epsilon/(35\pi)$, ${\kappa}^*=
36\,[1+\sqrt{1409}]\epsilon/385\pi$. The critical exponents for $d < 4$, to
lowest order in $\epsilon$, are $z=2-16\epsilon/35$, $\zeta = 1-2\epsilon/15$ 
(anisotropic scaling) and [since $u$ plays no role at $O(\epsilon)$] 
$\chi = 1-d/2$. These exponents clearly place this critical
point in a new universality class. 
A more detailed analysis, including the approach to the 
critical point, will appear elsewhere \cite{BIG}. 

We now investigate the low-temperature $v<0$ phase, {\em in the absence of noise}. 
It is convenient to
work  with dimensionless variables, obtained by  
rescaling $x_{\perp}$, $x_{\|}$, $t$ and ${\vs}$ in Eq.\
(\ref{eq:dyeq})\,; this leaves $\lambda$ as the only parameter 
in the equation of motion.
There are two static, spatially homogeneous steady states
--- a `paramagnetic
steady state' $\langle S_{\alpha} \rangle =0$, and a
`ferromagnetic steady state' 
$\langle S_1 \rangle = \langle
S_2\rangle = 0$ and $\langle S_3\rangle = 1$. 
It is straightforward to see from (\ref{eq:dyeq}) 
that both these stationary solutions are linearly unstable \cite{THESIS,BIG}. 
We next look for static, spatially
inhomogeneous steady states, a natural candidate being the
helical state. Defining $\rho \equiv \sqrt{S_1^2 + S_2^2}$ 
and $\phi \equiv \tan^{-1}(S_2/S_1)$,  
(\ref{eq:dyeq}) for $g_{\|}=g_{\perp}=0$ becomes  
\begin{eqnarray}
\frac{\partial \rho}{\partial t}&=& \nabla^2 \rho -\rho (\nabla \phi)^2
+\rho -(\rho^2+S^2_3)\rho -\lambda\rho S_3 \partial_{\|} \phi \, ,
\nonumber \\
\frac{\partial \phi}{\partial t}&=& \nabla^2\phi+\frac{2}{\rho}({\bf \nabla}
\rho)\cdot({\bf \nabla}\phi)+\frac{\lambda}{\rho}(S_3\partial_{\|}\rho
-\rho\partial_{\|}S_3) \, , \nonumber \\
\frac{\partial S_3}{\partial t}&=& \nabla^2 S_3 + S_3 -(\rho^2+S^2_3)S_3
+\lambda \rho^2 \partial_{\|}\phi \, .
\label{eq:cleq}
\end{eqnarray}
A regular helix $\rho=a$, $\phi=px_{\|}$ and $S_3=b$
($a,b$ and $p$ are arbitrary constants) is a steady state 
solution if $2 b^2 = 1-a^2(1+\lambda^2)\pm
\sqrt{(a^2(\lambda^2+1)-1)^2-4a^4}$\,, and $2 p= -\lambda b \pm \sqrt{
\lambda^2b^2-4(R^2-1)}$\, , where $R=\sqrt{a^2+b^2}$ is the magnitude of
each spin. The only free parameter $a$ is bounded by
$a<(3+\lambda^2)^{-1/2}$ from the requirement that $b$ be real.
Unfortunately even this steady state shows a linear instability, triggered by the growth
of $S_3$ \cite{THESIS}.

Having failed to find any stable static steady states analytically, we solve 
(\ref{eq:dyeq}) numerically for $d = 1$ without noise, for a range of generic initial 
conditions. 
To avoid numerical instabilities we adopt an
operator splitting method \cite{NUM} --- we solve the dissipative part
using the standard Euler method and the drive part 
\cite{BIG} by rotating each spin by an azimuthal angle
$\vert{\bf h}({x},t)\vert\triangle t$ about its computed local
magnetic field ${\bf h}$. With our choice of $\triangle x = 1$ and $\triangle t
=0.0001$ on a system of size $N=200$ with periodic boundary conditions, we
find that we avoid numerical instabilities and finite size effects.

\begin{figure}
\centerline{\epsfig{figure=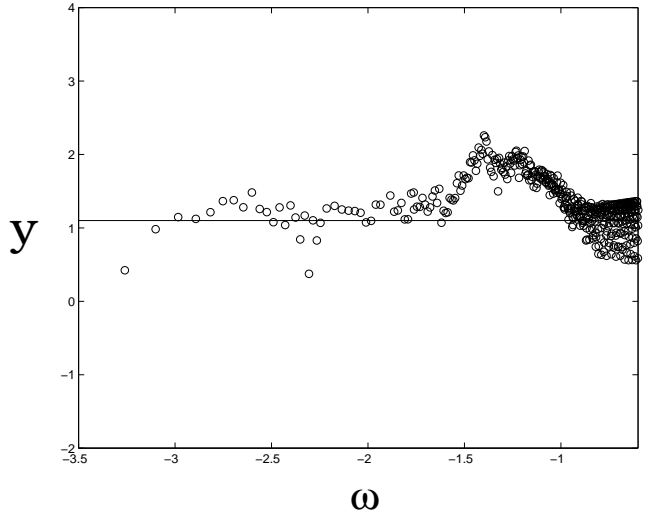,width=8.5cm,height=7.0cm}}
\caption{ Log-log plot of $y=\sqrt{\omega^2 \vert M_3(\omega)\vert^2}$ versus $\omega$
showing the $1/\omega^2$ dependence of the power spectrum over
approximately $1.5$ decades.}
\label{f1}
\end{figure} 

We find that the time series of the magnetization and
energy density $E = N^{-1}\int dx (\nabla {\vs})^2 $ never
settle to a
constant value; 
the motion could therefore be either temporally (quasi)periodic or chaotic. 
Figure.\ref{f1} shows that the power
spectrum of $M_3 \equiv \int_x S_3 $ goes as $1/\omega^2$.
The power spectrum of $E$ also shows a similar
behavior. This suggests that the dynamics is temporally chaotic
\cite{CROSS}. 

\begin{figure}
\centerline{\epsfig{figure=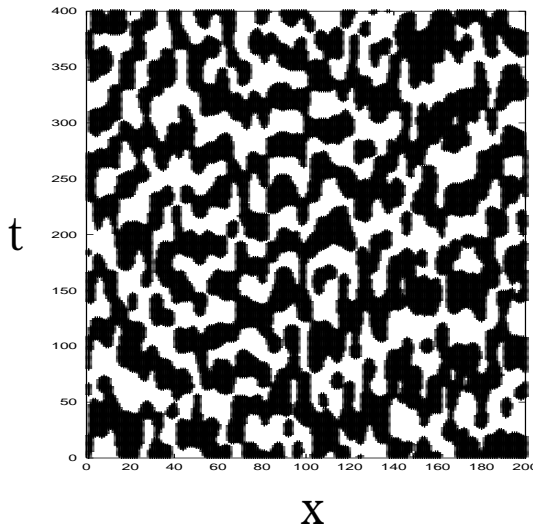,
width=7.0cm,height=7.0cm}}
\caption{ Space-time plot of the signed local pitch, 
sgn$(p)\equiv$ sgn$(\partial_x \phi)$ (black ($+1$), white ($-1$)),
revealing spatiotemporal chaos.}
\label{f2}
\end{figure}

Space-time plots of local quantities, such as the signed local pitch,
sgn$(p)\equiv$sgn$(\partial_x \phi)$ (fig.\ref{f2}), strongly suggest the
presence of spatiotemporal chaos\cite{CROSS}. These results are preliminary, and only
for $d = 1$. We shall characterize this behavior 
in greater detail elsewhere \cite{BIG}, including studies of the dependence of the
number of positive Lyapounov exponents on system
size and the behavior for $d > 1$. 

The helix solutions of Eq.  
(\ref{eq:cleq}) for $v < 0$ are an infinite family of unstable spatially periodic steady 
states (parametrised by $a$) of the type discussed in \cite{CHAOS}. Can chaos in our 
model be {\em controlled} so as to {\em stabilize} and {\it target} \cite{CHAOS} 
these helical states?
The control of spatiotemporal chaos in PDEs \cite{CHAOS,ASHWIN} is not nearly as
well-developed as that in finite dimensional dynamical systems \cite{CHAOS}.
Accordingly, it is significant that we are able to stabilize, target, and hence
control spatiotemporal chaos in our model, as we now show.  

For instance, in order to stabilize a specific helical configuration
(with fixed $a$, $b$ and $p$), we could in principle wait till the dynamics 
(presumably ergodic) 
eventually leads to this configuration, after which we apply small
perturbations to prevent $S_3$ from deviating from the value $b$ (recall that the
instability of the helical state was led by $S_3$). This
prescription successfully {\it stabilizes} the prescribed helix.

In order to {\it target} this prescribed helix, we add to (\ref{eq:cleq}) 
terms which would arise from a uniaxial spin anisotropy energy $V_3 = r_3 (S_3^2 -
b^2)^2$ or $V_3 = r_3 (S_3-b)^2$. We find that a sufficiently large and positive $r_3$  
forces $S_3$ to take the value $b$ exponentially fast starting from
arbitrary initial configurations. The subsequent evolution,  
given by Eq. (\ref{eq:cleq}) on setting $S_3=b$,
can be recast as purely relaxational dynamics,  
\begin{eqnarray}
\label{eq:downhill}
\frac{\partial \rho}{\partial t} = -\frac{\delta F}{\delta \rho}\,\,\, ,
\,\,\, \frac{\partial \phi}{\partial t} = -\frac{1}{\rho^2}
\frac{\delta F}{\delta \phi} \, ,
\end{eqnarray}
where the `free-energy functional' $F$ has the form of a {\it chiral
XY model},
\begin{eqnarray}
F  &  =  &  \frac{1}{2}\int_{\bf x} \left[(\nabla \rho)^2+\rho^2(\nabla
\phi)^2-(\rho^2+b^2)
+\frac{1}{2}(\rho^2+b^2)^2 \right. \nonumber \\
   &     &  \left. +\lambda b \rho^2 \partial_{\|} \phi \right] \,.
\label{eq:liap}
\end{eqnarray}
Using the chain-rule, it is easy to see that 
$dF/dt = \int_{\bf x} \left[- (\delta F/\delta S_1)^2
- (\delta F/\delta S_2)^2 \right] < 0$,
Hence $F$ is a Lyapunov functional for the dynamics. 
Completing the squares, we see that $\partial_{\|} \phi$ appears in $F$ in the
combination $(1/2)\rho^2(\partial_{\|} \phi + \lambda b /2)^2$, which is minimized by 
the helix $\phi = - (1/2)\lambda b x_{\|}$. 
Starting from any initial configuration, the system plummets towards
this unique helical minimum of $F$. 

Let us now see whether our control is robust against noise. We 
modify (\ref{eq:downhill}) by including the noise  
$\vec \eta$ in Eq. (\ref{eq:dyeq}), and ask for the statistics of small 
fluctuations with Fourier components $\tilde \rho_{\bf k}(t)$ and
$\tilde \phi_{\bf k}(t)$ about the controlled helical state, where 
$2 \pi/ L < k < \Lambda$ for a system of linear extent $L$. It is clear from
(\ref{eq:downhill}) that the means $\langle \tilde \rho_{\bf k} \rangle$ and
$\langle
\tilde \phi_{\bf k}\rangle$ decay exponentially to zero: the relaxation time
for $\tilde \rho_{\bf k}$ is finite at small $k$, whereas that for 
$\tilde \phi_{\bf k}$ goes as $k^{-2}$. To calculate the variances, 
note that the dynamics is governed in the mean by the 
Lyapounov functional (\ref{eq:liap}), and that the noise is spatiotemporally white. It 
follows \cite{CHAILUB} that the steady-state configuration 
probability $P[\rho, \phi] \propto e^{- c F}$ where $c$ is an effective inverse
temperature. $F \simeq \int[ {\rm const} (\tilde \rho)^2 + {\rm const} 
(\nabla \tilde \phi)^2]$ for small fluctuations about the helical minimum, i.e., $P$ 
is approximately gaussian, so that $\langle |\tilde \phi_{\bf k}|^2 \rangle \sim 
k^{-2}$ and $\langle |\tilde \rho_{\bf k}|^2 \rangle \sim {\rm const.}$ for small $k$. 
Thus the variance $\langle \tilde \rho^2 \rangle = 
\int_k \langle |\tilde \rho_{\bf k}|^2 \rangle$ is $L$-independent for $L \to \infty$ 
in any dimension $d$, whereas $\langle \tilde \phi^2 \rangle = 
\int_k \langle |\tilde \phi_{\bf k}|^2 \rangle$ diverges as $L$ and $\ln L$ respectively for $d = 1$ and
$2$, and is finite for $d > 2$. 
Thus {\it occasional excursions from the controlled state as a result of the noise 
do not lead to an instability of the targeted state for $d>2$}; the behavior for $d
\leq 2$ is no worse than for a thermal equilibrium $XY$ model.

Having arrived at the continuum equations Eq.\ (\ref{eq:dyeq}) based only on symmetry 
arguments and conservation laws, we now suggest ways in which 
the driving nonlinearity in (\ref{eq:dyeq}) may be realised.  
(a) Consider the isotropic magnet on a lattice whose unit cell lacks 
$\hat{\bf x} \to - \hat{\bf x}$ symmetry. Now subject the spins to a spatiotemporally 
random, isotropic, nonconserving noise source. The lack 
of invariance under $x$-inversion together with this nonequilibrium 
noise should give rise, via an interplay of nonequilibrium noise and the electrons 
participating in the exchange interaction, to the $\vec{S} \times \partial_x \vec{S}$ 
term. We admit, however, that we do not have a microscopic derivation of this. 
(b) Consider the asymmetric exclusion process (ASEP) \cite{DDLG} on a 1-dimensional 
lattice, each site $i$ of which can be either
vacant ($n_i(t) = 0$) or occupied by one particle ($n_i(t) = 1$) at time $t$. 
Each particle has {\em an attached 
Heisenberg spin} and may hop to the nearest neighbor at the right (left), if vacant,  
with probability $p$ ($q$). 
The spin $\vs_i(t)$ at an occupied 
site $i$ is the spin of the occupying particle.  
The local field at site $i$ is 
$\vh_i(t) = J_{i-1, i}\vs_{i-1}(t) + J_{i+1, i}\vs_{i+1}(t)$. 
The required asymmetric exchange is achieved by making the 
$J$s dynamical.  Recall that the exchange coupling between, say, $i-1$ and $i$ is 
operative at time $t$ only if $n_{i-1}(t) n_i(t) = 1$. Imagine that the local exchange
coupling emanating from a given site depended on some internal degree of freedom with a
short relaxation time, as follows: 
$J_{i, i \pm 1}(t) = J_1$ if $n_i(t-1) = 1$; if not, $J_{i, i \pm 1}(t) = J_2$. 
Assume for simplicity that the particles can hop only to the right.
Then a configuration $111$ at sites $i-1, \, i, \, {\rm and} \, i+1$ at time 
$t$ was either already present at time $t-1$ or arose from $011$ (by a right hop from 
$i-2$). Thus, for $111$ configurations, $J_{i-1,i}$ will be a weighted average of $J_1$
and $J_2$, while $J_{i+1,i}$ will be $J_1$. In the continuum limit, and averaging over 
the particle dynamics, we will get the driving term in (\ref{eq:dyeq}), with  
$\lambda \propto p - q$. Such averaging is justified if the hopping species has a faster 
dynamics than the spins (for example, a smaller dynamical exponent). Alternatively, 
allowing evaporation-deposition in the ASEP renders the ``particles'' fast 
without altering qualitatively the derivation above for the effective asymmetric 
exchange. Of course, particle non-conservation induces spin non-conservation trivially in 
this case. The work of \cite{radha} on moving space curves suggests 
another promising approach to finding realizations of our model.

In conclusion, we have studied the interplay of dissipation, precession, and spatially 
anisotropic driving on the dynamics of a classical Heisenberg magnet in $d$ space 
dimensions. We have found a nonequilibrium critical point which we have shown, in an 
expansion in $\epsilon = 4-d$, to be in a new dynamical universality class. We have 
presented evidence of spatiotemporal chaos in the mean-field dynamics of the model, at 
least in $d = 1$, and have shown how this chaos can be controlled to yield helical 
order. Our work reinforces the idea that spatiotemporal chaos is a generic feature of
driven, dissipative, spatially extended systems with nonlinear reactive terms.  
Further properties of this remarkable model, including the Lyapounov spectrum 
of the chaotic state, the possibility of complex ordered states or spatiotemporal chaos for $d>1$, possible
experimental realizations, and the
crossover between equilibrium and driven behavior, will be discussed elsewhere 
\cite{BIG}.  

\acknowledgments
We thank A. Dhar, Y. Hatwalne, B.S. Shastry, R.K.P. Zia, B. Schmittmann and 
U.C. T\"auber for discussions. JD is supported by NSF grant DMR-9727574.
MR thanks DST, India for a Swarnajayanthi grant.


\begin{thebibliography}{0}

\bibitem{MAMAZ} 
\Name{MA S. K. \and MAZENKO G. F.} \REVIEW{ Phys. Rev. B }{11}{1975}{4077}; 
\Name{FREY E. \and SCHWABL F.} \REVIEW{ Adv. Phys. }{43}{1984}{577}.

\bibitem{DDLG} \Name{SCHMITTMANN  B. \and  ZIA  R. K. P.}
\Book{Phase Transitions and Critical Phenomena}
\Editor{DOMB C. \and LEBOWITZ J. L.}
\Vol{17}
\Publ{Academic Press, NY}
\Year{1995}
\Page{1}.

\bibitem{HH}
\Name{HOHENBERG P. C.  \and  HALPERIN B. I.}\REVIEW{Rev. Mod. Phys.}{49}{1977}{435}.  

\bibitem{UWERUIZ} 
For other studies of driven multicomponent spin models, 
see, e.g., \Name{T\"AUBER U. C. , SANTOS J. E.  \and  R\'ACZ Z.} 
\REVIEW{Eur. Phys. J. B }{7}{1999}{309}, 
\Name{MARCULESCU S. \and  RUIZ RUIZ F.}\REVIEW{J. Phys. A }{31}{1998}{8355}
, \Name{T\"AUBER U. C., AKKINENI V. K. \and SANTOS J. E.}\REVIEW{Phys. 
Rev. Lett.}{88}{2002}{0457021} and \Name{ANTAL T., R\'ACZ Z. \and SASV\'ARI
L.}\REVIEW{Phys. Rev. Lett.}{78}{1997}{167}.

\bibitem{radha} 
\Name{BALAKRISHNAN R.}\REVIEW{J. Phys. C}{15}{1982}{L1305}. 


\bibitem{NOTE} If we add our $\lambda$ term to the usual \cite{MAMAZ} 
conserving dynamics for $\vec{S}$, standard perturbation theory 
at one-loop order yields non-conserving terms for the noise and propagator 
renormalisations at zero external wavenumber. 

\bibitem{THESIS}
DAS J.  in {\it Driven, Dissipative Dynamics of Heisenberg Spins with
Inertia}, Ph.D thesis (2000).  

\bibitem{BIG}
DAS J., RAO M. and RAMASWAMY S., in preparation.

\bibitem{KPZ} Thus, despite the superficial resemblance of 
this nonlinearity to that in the $d = 1$ 
stochastic Burgers equation \cite{DDLG}, there is 
no ``height'' representation [\Name{KARDAR M., PARISI G., \and ZHANG Y. C.}\REVIEW{Phys.
Rev. Lett.}{56}{1986}{889}] for eq.\,(\ref{eq:dyeq}).    


\bibitem{MA}
\Name{MA S. K.}\Book{Modern Theory of Critical Phenomena} 
\Publ{Benjamin, Reading, Mass}
\Year{1976}; 
see \Name{FORSTER D., NELSON D. R., \and STEPHEN M. J.}\REVIEW{Phys.
Rev. A}{16}{1977}{732}.

\bibitem{NUM}
\Name{PRESS W. H., TEUKOLSKY S. A., W. T. VETTERLING and FLANNERY 
B. P.}\Book{Numerical Recipes in Fortran, 2nd ed.}
\Publ{Cambridge University Press, Cambridge}
\Year{1992}.

\bibitem{CROSS}
\Name{CROSS M. C. \and HOHENBERG P. C.}\REVIEW{Rev. Mod. Phys.}{65}{1993}{851}.

\bibitem{CHAOS}
\Name{SHINBROT T., GREBOGI C., OTT E. \and  YORKE J. A.} \REVIEW{\it Nature}{363}
{1993}{411} ; \Name{SHINBROT T.}\REVIEW{Adv. Phys.}{44}{1995}{73}.

\bibitem{ASHWIN} 
\Name{SINHA S., PANDE A. \and PANDIT R.}\REVIEW{Phys. Rev. Lett.}{86}{2001}{3678}.

\bibitem{CHAILUB} 
\Name{CHAIKIN P. M. and LUBENSKY T. C.}\Book{Principles of Condensed
Matter Physics}
\Publ{Cambridge University Press, Cambridge} 
\Year{1995}.
\end{thebibliography}
\end{document}